\documentclass[12pt]{article}
\usepackage{graphicx}
\usepackage{color}
\usepackage{amsmath,slashed, amssymb}
\usepackage{fancyhdr}

\def\beq{\begin{eqnarray}}
\def\eeq{\end{eqnarray}}
\def\bea{\begin{eqnarray*}}
\def\eea{\end{eqnarray*}}
\def\bal{\begin{align}}
\def\eal{\end{align}}

\def\centeron#1#2{{\setbox0=\hbox{#1}\setbox1=\hbox{#2}\ifdim
\wd1>\wd0\kern.5\wd1\kern-.5\wd0\fi
\copy0\kern-.5\wd0\kern-.5\wd1\copy1\ifdim\wd0>\wd1
\kern.5\wd0\kern-.5\wd1\fi}}
\def\ltap{\;\centeron{\raise.35ex\hbox{$<$}}{\lower.65ex\hbox{$\sim$}}\;}
\def\gtap{\;\centeron{\raise.35ex\hbox{$>$}}{\lower.65ex\hbox{$\sim$}}\;}

\def\singleandthirdspaced{\baselineskip=\normalbaselineskip\multiply
    \baselineskip by 130\divide\baselineskip by 100}

\newcommand{\newc}{\newcommand}
\newc{\qbar}{{\overline q}}
\newc{\Kahler}{Kahler }
\newc{\deltaGS}{\delta_{\rm GS}}

\begin{document}
\begin{titlepage}
\begin{flushright}
{\large hep-th/yymmnnn \\
SCIPP 14/15\\
}
\end{flushright}

\vskip 1.2cm

\begin{center}

{\LARGE\bf Naturalness Under Stress}

\vskip 1.4cm

{\large Michael Dine}
\\
\vskip 0.4cm
{\it Santa Cruz Institute for Particle Physics and
\\ Department of Physics,
     Santa Cruz CA 95064  } \\

\vskip 4pt

\vskip 1.5cm

\begin{abstract}
Naturalness has for many years been a guiding principle in the search for physics beyond the Standard Model, particularly for understanding the physics of electroweak symmetry breaking.  However, the discovery of the Higgs particle at 125 GeV, accompanied by exclusion of many types of new physics expected in natural models has called the principle into question.   In addition, apart from the scale of weak interactions, there are other quantities in nature which appear unnaturally small and for which we have no proposal for a natural explanation.We first review the principle, and then discuss some of the conjectures which it has spawned.  We then turn to some of the challenges to the naturalness idea and consider alternatives.  
\end{abstract}

\end{center}

\vskip 1.0 cm

\end{titlepage}
\setcounter{footnote}{0} \setcounter{page}{2}
\setcounter{section}{0} \setcounter{subsection}{0}
\setcounter{subsubsection}{0}

\singleandthirdspaced

\section{Naturalness:  A Contemporary Implementation of Dimensional Analysis}

In our first science courses we learn about the importance of dimensional analysis.  Often this is presented as a consistency requirement for calculations of physical quantities.  But it shapes our understanding of physical systems in a fundamental way. For example, from the electron mass, $m_e$,  the speed of light, $c$, and Plancks constant, $\hbar$ we can form a quantity with dimensions of length:
$
a = {\hbar \over m c} \approx 10^{-15} {\rm cm}.
$
supplemented with the insight that the strength of the force between the proton and the electron is proportional to $e^2$, so the size should get larger as $e^2$, or
$
a = {\hbar \over m c  e^2}.
$
To know the exact coefficient -- which is an order one number -- we need to solve the Schrodinger equation completely.  But we get a nice qualitative, and rough quantitative, picture without much trouble.

Similarly, the size of atomic nuclei is large compared to the Compton wavelength of the proton.  This in turn suggests there should be physics associated with this larger length scale.  Without worrying about the detailed mechanism, this suggests the existence of a particle with a mass roughly equal to that of the pion.  


This sort of reasoning has successes in many other areas of physics.  What is more interesting is questions where it fails, at least at first sight. In 1899 Planck noted that from $\hbar, ~c$, and $G_N$, one can form a quantity with units of mass, $
M_p = \sqrt{c \hbar/G_N}.
$
At or below this scale, quantum mechanical effects should be important
in general relativity.

Suppose that there is some underlying theory from which one compute the electron mass, which includes general relativity.  Dimensional analysis would say that 
$
m_e = \beta~M_p
$
where $\beta$ is an order one number.  Of course, this is terribly wrong -- dimensional analysis fails stupendously here.

Lorentz encountered this issue in a somewhat different way, which provides a different -- and equally useful -- perspective on the problem.  Lorentz modeled the electron as a smooth charge distribution with a characteristic size, $a$.  One would expect that the mass of the electron would then be at least of order
the self energy of the electron arising from its Coulomb field,
$
m_e \approx {e^2 \over 4 \pi a}.
$
This might be viewed as a prediction of $a$:  $a \approx 10^{-10}$ cm or even $10^{-12}$ cm.  But from present day experiments, we know that $a < 10^{-17}$ cm.  This is then, at first sight, a serious failure of dimensional analysis.     Alternatively, we might describe this as a problem of ``naturalness", or {\it fine tuning}.   If there is an additional, ``bare", mass parameter, $m_e^{(0)}$,
$
m_e = m_{e}^{(0)} + {e^2 \over a}
$  Each term separately is about $5 \times 10^4$ the observed mass of the electron.

The resolution of this puzzle has been known since the work of Weisskopf in 1934\cite{weisskopf1,weisskopf2}.  His supervisor at the time, Wolfgang Pauli, assigned him the problem of computing the corrections to the energy of a free electron due to ints interactions with its own fields.  Using the newly discovered rules of (relativistic) quantum mechanics, this required including not only the interaction of the electron with its Coulomb field, but contributions to the energy from intermediate states of two types, one with an electron and a photon, and one with two electrons, a positron, and a virtual photon.  The expressions were divergent at high energies (corresponding, in modern language, to high virtual photon momenta), and Weisskopf assumed that these were cut off by the size of the electron.  In his first attempt, he encountered a similar linear divergence ($1/a$) as in Lorentz's calculation, but, following an observation of Furry, he quickly corrected a mistake and found that the leading linear divergence cancelled, leaving only a logarithmic dependence on the cutoff.  The full expression, which can be derived by a modern field theory student in a matter of minutes, is
\beq
m_e =  m_{e}^{(0)} \left (1 -  {6\alpha \over 4 \pi} \log(m_e a)\right ).
\label{electronmass}
\eeq
Even for extremely small $a$, $a = 10^{-31} ~{\rm cm}$, the correction is only about $20\%$ of the leading result.
It is remarkable that the ``naturalness" problem of the classical theory is resolved, not simply by the quantization of the theory, but by the fact that there are additional degrees of freedom required by the relativistic quantum theory.  In fact, if the electron had been a scalar, this would not have happened; as we will discuss further below; instead the mass-squared diverges quadratically with the cutoff.

It is crucial that eqn. \ref{electronmass} is proportional to the original electron mass, the parameter which appears in the lagrangian for the theory.  This can be understood in a conceptual way.  In the limit that the mass of the electron vanishes, quantum electrodynamics is more symmetric.  Setting the mass term, $m_e^{(0)}$, in the usual Dirac lagrangian,
\beq
{\cal L} = \bar \psi \left ( i \gamma^\mu (\partial_\mu - i e A_\mu) -  m_{e}^{(0)} \right ) \psi
\eeq
to zero, one has a symmetry under the {\it chiral} or {\it Weyl} transformation:
$
\psi \rightarrow e^{i \omega \gamma_5} \psi.
$
In this limit, all effects in the theory -- and in particular any corrections to the lagrangian -- must respect the symmetry.  This means, in particular, that any correction to the mass must vanish as the mass tends to zero, precisely the feature of eqn. \ref{electronmass}.

So we see that, while even now we don't have a compelling microscopic explanation of this mass, small $m_e$ is special in that
Quantum Electrodynamics becomes more symmetric.  't Hooft elevated this to a {\it principle of naturalness}:  a quantity in nature should be small only if the underlying theory becomes
more symmetric as that quantity tends to zero\cite{thooftnaturalness}.  
There are other instances where this reasoning works remarkably well.  Consider, for example, the mass of the proton.  The proton is composed of quarks and gluons, but its mass has very little to do with the masses of the quarks, which is of order the small difference between the proton and neutron masses.  So again, we might ask why the mass of the proton is not $M_p$.  The answer turns out, again, to be related to symmetries.  Setting the quark masses to zero, the classical action of QCD has no scale -- the theory has a symmetry called {\it scale} or {\it conformal} invariance.  If this symmetry were exact, the proton would necessarily be massless; in the quantum theory this symmetry is broken by a small amount.  

The violations of scale invariance are associated with the phenomenon of renormalization in quantum field theory.  Renormalization is the statement that the parameters of a theory vary with length or energy scale.  This variation is logarithmic, encoded in {\it renormalization group equations}.  For the strong coupling, $\alpha_s$, specifically:
\beq
{d \alpha_s \over dt} = -2~b_0 \alpha^2.
\eeq
Here $t = \log(M/E)$, where $M$ is an ultraviolet cutoff (or matching scale), and $b_0$ a constant.  So if one asks at what
scale $E \equiv \Lambda$, the coupling becomes of order one:
\beq
\Lambda = M_p e^{-{2 \pi \over b_0 \alpha_s(M_p)}} 
\eeq
For QCD, $b_0$ is a number of order $7$, so if $g_s$ at $M_p$ is about  $0.5$, the exponential is extremely small, and  the scale $\Lambda$ is of order the proton mass.

Most of the parameters of the Standard Model (SM) are natural in the sense of 't Hooft.  But there are some quantities which are not.  It is precisely the failures of dimensional analysis which are most interesting.  As for the electron and proton masses, they have the potential to point to possible new phenomena in nature -- new degrees of freedom, interactions and/or symmetries.  For a long time, these sorts of puzzles have served as a guide to speculation about physics beyond the Standard Model.




\section{Naturalness Problems in Particle Physics}

Our current theories of the laws of nature are best viewed as tentative, {\it effective} field theories, valid at energies below some scale at which new degrees of freedom or other phenomena might manifest themselves.  Naturalness, from this perspective, is the assertion that features of this effective
field theory should not be extremely sensitive to the structure of the underlying theory.  For the electron,
this is the statement that its Yukawa coupling receives only small corrections as one studies the theory at progressively higher energy scales.
For the strong interactions, as we have seen, this is the statement that the existence of a proton much lighter than the Planck 
scale can be explained by an ${\cal O}(1)$ pure number at $M_p$.  

The masses of the quarks and leptons are controlled by symmetries much as the mass of the electron in the Weisskopf calculation.
First, the $SU(2) \times U(1)$ symmetry of the SM forbids masses smaller than
$
 y \times 250 ~{\rm GeV}.
$
Here $y$ is a pure number, the Yukawa coupling of the quark or lepton.  For the quarks and charged leptons, this number ranges from about $1$ for the top quark to $10^{-5}$ for the electron.  The spread in these numbers raises many puzzles, but it is not unnatural.  Just as was the case of the small electron mass in pure quantum electrodynamics, in the limit of very small electron Yukawa coupling, the theory becomes more symmetric.  Indeed, if we set all of the Yukawa couplings to zero, the theory possesses a large symmetry.   A number of theories have been proposed which might account for these small numbers and the hierarchies among them.  It is fair to say that none is completely compelling by itself, nor do any make unequivocal predictions for experiment.  Still, the existence of a hierarchy in fermion masses and mixings does not pose a fundamental conceptual problem.

There is one quantity in the SM which fails 't Hooft's test and raises precisely the sorts of issues posed by the classical theory of the electron.  This is the mass of the Higgs particle, which is tied to the scale at which the symmetry of the electroweak theory is broken. In the simplest version of the SM, the potential of the Higgs field is
\beq
V(\phi) = -\mu^2 \vert \phi \vert^2 + {\lambda \over 4} \vert \phi \vert^4.
\eeq
Assuming that this potential describes the recently observed Higgs particle (and measurements to date are consistent with this picture), we know the values of $\mu$ and $\lambda$:
$
\mu \approx 89 ~{\rm GeV};~\lambda \approx 0.13.
$

Dimensional analysis, on the other hand, would predict $\mu^2 \approx M_p^2$, and there is no enhancement of the symmetry of the theory if we take $\mu^2 \rightarrow 0$.  If we repeat Weisskopf's calculation for this case, we confront this issue directly.
The strongest coupling of the Higgs field in the SM is its
Yukawa coupling to the top quark:
$
{\cal L}_{\bar t t H} = y_t H Q_3 \bar t
$
where $Q_3$ refers to the third quark doublet, consisting principally of the $t$ and $b$ quarks.
\begin{figure}
\includegraphics[width=8cm]{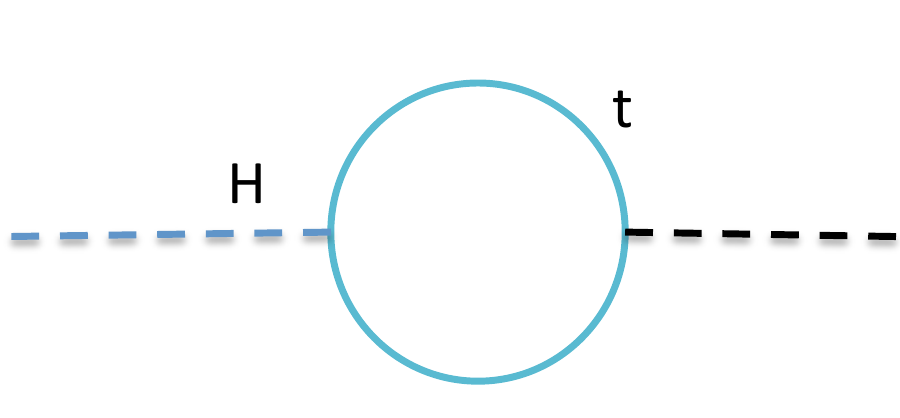}
\caption{One loop correction to Higgs mass involving top quarks.}
\end{figure}
At one loop, there is a correction to the Higgs mass coming from the diagram of fig. 1.  This is given by:
\beq
\delta \mu^2 = -6 y_t^2 \int {d^4 k \over (2 \pi)^4} {1 \over k^2}
\eeq
where the integral is an ordinary Euclidean integral.  This diverges quadratically.  If the cutoff is the Planck scale, this correction is enormous, consistent with expectations of dimensional analysis, about thirty four orders of magnitude larger than $M_H^2$, corresponding to a fine tuning of the bare parameters against the radiative correction at the part in $10^{34}$ level.

This is one of many such contributions to the Higgs mass, including contributions from diagrams involving lighter quarks, gauge bosons and the Higgs themselves.
The issue of the quadratic growth (divergence) in corrections to scalar masses was first raised by Ken Wilson\cite{wilsonhierarchy}.

\section{Other ``Unnatural" Standard Model Parameters}


We have remarked that the small quark and lepton masses (Yukawa couplings) are natural in the sense of 't Hooft.
In the limit that all of the quark and lepton masses vanish, the Standard Model has a large global symmetry,  For each type of quark or lepton (where type is defined the gauge quantum numbers of the associated field)
$
Q_f, \bar U_f, \bar d_f, L_f, \bar e_f
$ 
the theory has a separate $U(3)$ symmetry, defined, in the case of $Q_f$, for example, by
$
Q_f \rightarrow U_{f,f^\prime} Q_f^\prime.
$
As a result, quantum corrections to the Yukawa couplings (and hence masses) vanish in the limit that the masses tend to zero.\footnote{There is an exception associated with the fact that one linear combination of the $U(1)$ subgroups of these symmetries has a QCD anomaly.  However, Feynman diagram corrections still vanish, and the resulting effects are quite small.}
Many physicists have explored the possibility that some underlying theory possesses precisely these symmetries (or perhaps a continuous or discrete subgroup), and that they are spontaneously broken by a small amount.

There is one small parameter which does appear, on its face, to violate 't Hooft's condition.  It is possible to add to the QCD lagrangian a term
\beq
{\cal L}_{\theta} = {\theta \over 16 \pi^2} G_{\mu \nu} \tilde G^{\mu \nu}.
\eeq
Here $G_{\mu \nu}$ is the QCD field strength, and $\tilde G_{\mu \nu}$ 
is its dual:
$
\tilde G_{\mu \nu} = {1 \over 2} \epsilon_{\mu \nu \rho \sigma} G^{\rho \sigma}.
$
This term is odd under parity ($P$) and even under charge conjugation ($C$), so it violates CP.
In electrodynamics, the analogous term is $\vec E \cdot \vec B$, which is a total derivative, and has no effect.\footnote{This is not quite true; if there are magnetic monopoles in nature,a $\theta_{qed}$ parameter effect the properties of their charged excitations, the ``dyons"\cite{witteneffect}.}  In QCD, the term is also a total derivative.  As a result, it does not affect the equations of motion.  However, it does have physical effects.  Most notably, using current algebra one can can compute the electric dipole moment of the neutron, $d_n$, as a function of $\theta$\cite{Crewther:1979pi}:
\beq
d_n
= 5.2 \times 10^{-16} \theta ~{\rm cm}    .
\eeq
From the experimental limit, $d_n < 3 \times 10^{-26} ~{\rm e~cm}$, one has
$\theta < 10^{-10}.$
If nature respected $CP$ in the absence of $\theta$, this small value of a dimensionless number would be natural in the sense of 't Hooft.  But nature violates CP; indeed, the phase appearing in the CKM matrix is of order one.  So, like the Higgs mass, this number cries out for an explanation.

\section{Proposed Solutions to the Problem of the Higgs Mass}

Over the years, several solutions of the hierarchy problem have been proposed.

\subsection{Technicolor}

Susskind and Weinberg put forward the first solution to the problem of naturalness of the Higgs mass, closely paralleling the understanding of the hierarchy between the proton mass and the Planck scale
\cite{weinberghypercolor,susskindtechnicolor}.  They proposed that electroweak symmetry breaking arises due to a condensate of fermions
in some new strong interactions, similar to QCD but with a scale of order $1 ~TeV$.  Susskind dubbed this solution {\it technicolor}.

 Consider the SM without the Higgs particle, and with only a single generation of quarks and leptons, i.e. with fermions:
 \beq
Q = \left ( \begin{matrix} u \cr d \end{matrix} \right );~~\bar u~~~\bar d;~~L = \left ( \begin{matrix} \nu \cr e \end{matrix} \right );~~\bar e.
\eeq
Neglecting thel weak coupling, the quark sector of the theory  possesses a global symmetry
$SU(2)_L \times SU(2)_R \times U(1) \times U(1)$.  $SU(2)_L$ is just the $SU(2)$ of weak interactions, which rotates the doublet $Q$; $SU(2)_R$ is an approximate symmetry under which $\bar u$ and $\bar d$ transform as a doublet.  The $U(1)$ of the SM is a combination of the diagonal generator of the $SU(2)_R$ as well as one of these $U(1)$'s.  The strong interactions break the symmetry to the diagonal subgroup, the familiar $SU(2)$ of isospin, as well as a $U(1)$; a linear combination of these symmetries is electric charge.

Because the $SU(2)_L \times U(1)$ subgroup of this symmetry is gauged, the $W$ and $Z$ gain mass, and the photon remains massless.  This
is nicely illustrated using the familiar non-linear lagrangian description of chiral symmetry breaking, where the pions are described by a matrix of fields with a simple transformation property under the $SU(2)_L \times SU(2)_R$:
\beq
\Sigma = e^{i {\pi^a \sigma^a \over 2 f_\pi}};~~~\Sigma \rightarrow U_L \Sigma U_R
\eeq
The lagrangian for $\Sigma$ is:
\beq
{\cal L}_{\Sigma} = f_\pi^2 {\rm Tr} \left ( D_\mu \Sigma D_\mu \Sigma^\dagger \right ).
\eeq
It is an instructive exercise to work out the form of the covariant derivatives (the reader for whom this is not familiar would do well to first do the exercise of just gauging the $SU(2)_L$, where the gauge interactions only act from the left; then include the $U(1)$ by gauging a subgroup of the $SU(2)_R$).  With this, one immediately finds that the gauge boson masses are just those of the SM, with the Higgs expectation value, $v$, replaced by $f_{\pi}$.

The technicolor hypothesis just replaces the ordinary quarks by techniquarks, and color by a new interaction, $f_\pi \rightarrow F_{TC} = v$.  This theory solves the hierarchy problem both in the sense that there are no longer quadratic divergences (loosely the divergences are cut off at the technicolor scale), and also in that it provides an explanation of the weak scale, analogous to the QCD explanation of the proton mass:
$
F_{tc} = M e^{-{8 \pi^2 \over b_{tc} g_{tc}(M)^2}}.
$

While a beautiful idea, this proposal runs into a number of difficulties.  First, in this simple form, it has no mechanism to account for the masses of quarks and leptons.
One can try to resolve this problem by introducing further gauge interactions, whose role is to break the chiral symmetries which protect fermion masses.  The resulting models are quite baroque, requiring many gauge groups and intricate dynamics, but aesthetic objections aside, they run into serious issues with flavor changing neutral current processes.  Put simply, the Standard Model possesses a variety of approximate symmetries due to small quark masses, and these account, for example, for the small rate for $K \leftrightarrow \bar K$ mixing; it is difficult to mimic
this phenomenon in a strongly interacting theory.

Prior to the Higgs discovery, other serious problems have long been noted, especially difficulties with precision studies of the Standard Model\cite{peskintakeuchi}.  The existence of a Higgs much lighter than $1$ TeV, and with width less than a few GeV,
is particularly difficult to understand in a Technicolor framework.  Most proposals to understand this assume that the technicolor theory
is nearly conformal over a range of scales, with a light, SM-like Higgs a consequence.



\subsection{Little Higgs and Similar Models}

An approach which attempts to reconcile the idea of dynamical electroweak symmetry breaking with the existence of a Higgs particle
light compared to the scale of interactions is to consider the Higgs as an approximate Goldstone boson\cite{littlehiggs1,littlehiggs2,littlehiggs3}.  The basic idea of such {\it Little Higgs Models} is that there are some new strong interactions, at a scale $M$, and that these interactions possess an approximate global symmetry which is spontaneously broken.  One of the Goldstone bosons of this symmetry acts as the Higgs boson.  The Standard Model gauge interactions necessarily break these symmetries and give rise to a potential.  The Higgs mass term induced is too large unless one introduces additional features in such a way that the approximate
symmetries are violated only by combinations of additional gauge symmetries.  Accounting for fermion masses and satisfying other constraints
is challenging.

At a perhaps more drastic level, ref. \cite{grinsteinwise} suggests a modification of the conventional structure of quantum field theory, through "Lee-Wick" theories.  Assuming this is
the solution of the hierarchy problem, one again predicts new physics at the TeV scale.

\subsection{Large Extra Dimensions}

In the large extra dimension models\cite{largeextradimensions1,largeextradimensions2}, one alters the nature of the hierarchy problem by postulating that the fundamental scale of physics is near the scale of electroweak breaking, of order $TeV$.  This can be accommodated if one supposes that there are some number, $d$, of compact extra dimensions of space (minimally two), with volume $\ell^d$.  Then starting with the $d+4$ dimensional Einstein action,
\beq
{\cal L}_{d+4} = \kappa^{-2}_{d+4}\int d^4 x d^d y \sqrt{g} {R}
\eeq
where $\kappa^2$ is the $d+4$ dimensional Newton constant and $y$ are the extra dimensions, the four dimensional Newton constant is simply
$G_N = {\kappa^2 \over \ell^d}$.
If
$
\kappa^2 = ({\rm TeV})^{-(2+d)}, then
$
for $d=2$, for example, the dimensions are of order 
millimeters; for $d=6$, one has $\ell \approx 0.2~{\rm MeV})^{-1} \approx 10^{-9} ~{\rm cm}$.  

In order that one not have a similar dilution of the strength of the Standard Model interactions by $\ell^d$, these theories need an additional feature:  the Standard Model must live on a geometric object known as a 3-brane. {\it P-branes} are generalizations of membranes ($2$-branes), strings ($1$-branes) and particles ($0$ branes).  These 3-branes fill all of space; excitations on the $3$-brane behave like particles in four dimensions.  The Standard Model gauge bosons, fermions, and Higgs boson, in this picture, are excitations of the brane. 

These models make exciting predictions\cite{largeextradimensionphenomenology}.  In the two dimensional case, for example, one predicts modification of Newton's laws at millimeter scales, and this has prompted experimental searches which have constrained the possibility by verifying Newton's laws to remarkably small distances\cite{adelberger1,adelberger2}.  Such models also predict the existence of many new particles, associated with the modes of the higher dimensional fields on the compact volume (Kaluza-Klein modes).  At sufficiently high energies one should produce large numbers of these particles, essentially uncovering the physics of the higher dimensional space time.

This approach alters the question of hierarchy to the question:  why are these extra dimensions so large?  It remains to find a compelling picture, but the possible is intriguing and possible short distance modifications of GR or signals of large extra dimensions in accelerators remain active subjects of investigation.

\subsection{Warped Extra Dimensions}

Warped extra dimensions incorporate elements of the large extra dimension picture and of technicolor models\cite{rs1,rs2}.  Here, one also has extra dimensions (for simplicity we will consider one extra dimension) and 3-branes.  In a simple version, the Standard Model sits on one of two branes.  Under certain conditions, the Einstein equations in the higher dimensional space admit a solution where the metric varies exponentially with the distance from one or the other brane.   This variation is analogous to the variation of couplings with scale in non-abelian gauge theories that we have encountered earlier.  The strength of gravity relative to the weak interactions is exponential in the separation of the branes, $e^{-\ell}$.   As a result, gravity is very weakly coupled on one brane, and strongly coupled on the other.  Variants of this idea have some of the Standard Model fields in the {\it bulk} space between the branes.  Scenarios exist which would account for quark and lepton masses and the suppression of flavor changing processes. 
Precision electroweak physics and the observed Higgs particle pose significant challenges, as does embedding this picture in
a more complete theory such as string theory.
Signals of such warped dimensions include low lying Kaluza-Klein states, and a great deal of effort has gone into searching for such particles.

\section{Supersymmetry}


In implementing 't Hooft's notion of naturalness, we have so far considered symmetries of a sort familiar from quantum mechanics, generated by a charge operator which is a scalar under rotations.  But there is another type of symmetry, allowed by general principles of quantum mechanics and relativity, where the symmetry generators are {\it spinors}.  This symmetry is known as supersymmetry.   We will consider it, first, as a global symmetry, but the symmetry can be elevated to a local, gauge symmetry.

Supersymmetry has many remarkable properties.  First is the algebra of the symmetry generators; these obey anti-commutation relations with the energy and momentum on the right hand side:
\beq
\{Q_\alpha, Q^*_{\dot \beta}\} = P_\mu \sigma^\mu_{\alpha \dot \beta}.
\label{susyalgebra}
\eeq
Here $P^\mu$ is the total four momentum of the system.  We are using two component spinor notation, with
$\sigma^i_{\alpha \dot \beta}$ the ordinary Pauli matrices, while $\sigma^0$ is the identity matrix.
Taking the trace of both sides gives:
\beq
\sum_\alpha Q_\alpha Q_\alpha^* + Q_\alpha^* Q_\alpha = 2 E.
\label{anticommutator}
\eeq
As for any symmetry, these generators (charges) commute with the Hamiltonian.  Acting on bosonic or fermionic states one has relations:
\beq
Q_\alpha \vert B \rangle \propto \sqrt{E} \vert F \rangle~~~~~ Q_\alpha \vert F \rangle \propto \sqrt{E} \vert B \rangle
\eeq
As a result, if the symmetry is exact and unbroken, fermions and bosons are degenerate.  

This feature of supersymmetry makes it particularly interesting for the hierarchy problem.    Among the bosons
of supersymmetric theories are fundamental scalars.  We have seen that
fundamental scalars provide a very simple way to understand quark and lepton masses; they have the further advantage that they are consistent with precision electroweak studies and now the discovery of what appears to be an elementary Higgs scalar.  So it would be desirable to find theories in which the masses of elementary scalar fields were protected by symmetries.  Supersymmetry is the only known such symmetry.
As we have explained, it is natural for fermions to be light; in the presence of supersymmetry, it follows that it is also natural for bosons, and in particular scalars, to be light.

Of course, there is no such degeneracy among the particles of the Standard Model, so the symmetry must be broken; to account for the Higgs
mass, in the spirit of 't Hooft's principle, the breaking scale should be much smaller than
the Planck scale.  Witten pointed out some time ago that supersymmetry is particular susceptible to small, spontaneous breaking\cite{Witten:1981nf}.   From eqn. \ref{anticommutator} supersymmetry is unbroken if and only if $E=0$.  It turns out that supersymmetric field theories for which $E=0$ classically have $E=0$ (and unbroken supersymmetry) to all orders in perturbation theory\cite{nonrenormalization}.  But this need not hold beyond perturbation theory, and often does not.  This means that the energy scale of supersymmetry breaking can take the form:
$
E = M e^{-{8 \pi^2 \over g^2}}
$
reminiscent of other hierarchies we have encountered. This phenomenon is referred to as {\it dynamical supersymmetry breaking}. 

\subsection{Basics of Supersymmetric Field Theories}

There are a variety of excellent texts and review articles on supersymmetry.   There is not space here to
fully elucidate the structure of supersymmetric theories, but a few basic features will be helpful for our subsequent discussion.
\begin{enumerate}
\item  Supersymmetry multiplets:  In globally supersymmetric models, there are two basic types of multiplets:  chiral multiplets, consisting of a complex scalar and a spin-1/2 fermion, and vector multiplets, consisting of a chiral fermion and a gauge boson.
\item  Interactions of the matter fields with each other:  These are described by a holomorphic (analytic) function of the chiral fields (scalar components) called the superpotential, $W(\phi_i)$.  
In terms of $W$, there is a contribution to the potential for the scalars:
\beq
V_W = \left \vert {\partial W \over \partial \phi_i} \right \vert^2
\eeq
as well as mass terms and Yukawa couplings for the fermions:
\beq
{\cal L}_f = {1 \over 2} {\partial^2 W \over \partial \phi_i \partial \phi_j} \psi_i \psi_j + {\rm c.c.}
\eeq
where the $\psi_i$'s are the fermionic partners of the $\phi_i$'s.  
\item  Interactions of the gauge fields with each other:  in a particular gauge (Wess-Zumino gauge), the vector fields interact with each other just as in ordinary non-abelian gauge theories; the gauginos, $\lambda^a$, couple to the gauge fields as expected for fermions in the adjoint representation.
\item  Interactions of the matter fields and the gauge fields:  in the same gauge, the scalars and fermions in the chiral multiplets couple to gauge fields just as in ordinary gauge theories.  They possess Yukawa couplings to the gauginos:
\beq
{\cal L}_{\lambda} = \sqrt{2} g (\lambda^a \phi_i^* T^a \psi_i) + {\rm c.c.}
\eeq 
\item  Quartic couplings of scalars charged under the gauge groups:
\beq
V = {g^2 \over 2} (\phi^*_i T^a \phi_i)^2.
\eeq 
\end{enumerate}

Supersymmetry can be elevated to a local symmetry.  In that case, the gauge field associated with {\it local supersymmetry transformations} is the gravitino, $\psi^\mu_\alpha(x)$, a field of spin 3/2.  The action becomes distinctly more complicated\cite{supergravitya,supergravityb}.  In the limit of unbroken supersymmetry in flat space, one can define global supercharges, just
as one can define a global energy and momentum.  These global supercharges still obey the basic algebra, eqn. \ref{susyalgebra}.  In addition to the chiral
and vector multiplets, there is a gravitational multiplet, consisting of the graviton and a spin-3/2 fermion, the gravitino.
Small breaking of supersymmetry in supergravity leads to theories which look, at low energies, like globally supersymmetric theories with explicit soft breaking\cite{giradello}. 

\subsection{Building Models for Supersymmetry and Its Breaking}

If nature is supersymmetric, the partners of the  known fermions (quarks and leptons) are complex scalar fields (with the same gauge charges).  These particles are referred to as squarks and sleptons.   The partners of the gauge bosons are the gauginos.    The fermionic partners of the Higgs fields (supersymmetry requires a minimum of two Higgs doublets) are known as {\it higgsinos}.

Constructing realistic models with dynamical supersymmetry breaking poses challenges, so most searches for supersymmetry, and many investigations of  the basic features of such theories, start by introducing an explicit, soft breaking of the symmetry,  This amounts to simply adding masses for the squarks, sleptons and gauginos, as well as certain dimensionful couplings\cite{Dimopoulos:1981zb}.  These parameters (along with cubic couplings of the scalars) are called soft because they do not spoil the good ultraviolet properties of the theories\cite{giradello}.

\begin{figure}

\includegraphics[width=8cm]{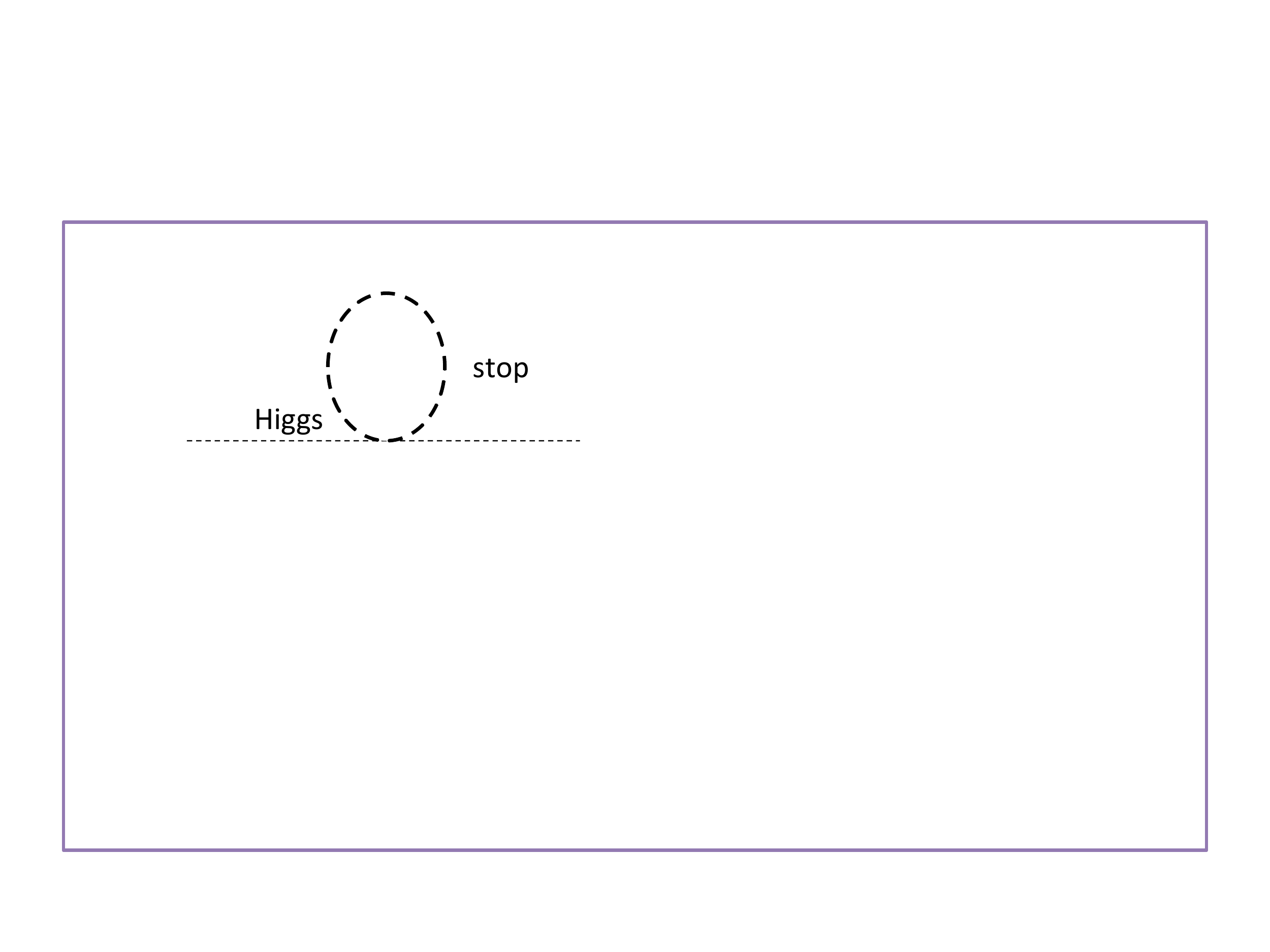}
\caption{Additional correction to Higgs mass from stops.}
\end{figure}
 In addition to the top quark loop which we have discussed previously, there is now a loop (fig. 2) containing a stop which tames
 the quadratic divergence of the SM.  There are actually two types of stops, one from an electroweak doublet, one from the singlet.  For simplicity, calling the mass of each of these scalars $\tilde m_t^2$, the two Feynman diagrams yield:
\beq
\delta m_H^2 = 3 y_t^2 \int {d^4 k\over (2 \pi)^4} \left ( -{1 \over k^2 + m_t^2} + {1 \over k^2 + \tilde m_t^2} \right ).
\eeq
The minus sign in the first term is the usual minus sign in field theory associated with fermion loops.  The leading quadratic divergence cancels, leaving only a logarithmically divergent term:
\beq
\delta m_H^2 = -{3 y_t^2 \over 16 \pi^2} \tilde m_t^2 \log(\Lambda^2/\tilde m_t^2).
\label{stopcontribution}
\eeq
Here $\Lambda$ is an ultraviolet cutoff, and we have assumed $m_t^2 \ll \tilde m_t^2$, consistent with exclusions from the 
Large Hadron Collider (LHC), which we will discuss shortly.  This is closely parallel to the situation for the electron mass in QED.  

\section{The simplest implementation of supersymmetry:  the MSSM}

To develop a supersymmetric phenomenology, we can promote each fermion of the SM to a chiral multiplet and each
gauge boson to a vector multiplet.  We have quark doublets and antiquark singlets $(Q_f, \bar u_f, \bar d_f)$, and
lepton doublets and singlets ($L_f, \bar E_f$), with $f$ a flavor label.There are necessarily two Higgs doublets, $H_U$ and $H_D$ (otherwise the model is inconsistent), and gauginos accompanying each of the gauge bosons. 

The superpotential of the model includes couplings of the Higgs to quarks and leptons:
\beq
W = y^D_{fg} Q_f \bar D_g H_D + y^U_{fg} Q_f \bar U_g H_U + y^L_{f} L_f \bar \bar E_f H_D.
\eeq
The expectation value of the Higgs accounts
for the fermion masses.
There are, in addition, a variety of renormalizable terms which can lead to processes in which baryon and/or lepton number are violated.  Terms in the superpotential such as $
Q L \bar \tilde d ~~~{\rm and}~~~ \bar u \bar d \bar d$
lead to violation of baryon and lepton numbers.  If the dimensionless coefficients are of order one, one would expect the proton to decay in about $10^{-24}$ seconds.  Postulating a discrete symmetry, called $R$ parity, forbids these operators.   Under this symmetry,  all of the particles of the SM, as well as the additional Higgs doublets, are even, while all of their superpartners are odd.  With this restriction, the MSSM contains $105$ new parameters, associated with the soft breaking of supersymmetry and the
additional Higgs field.  Consistent with 't Hooft's principle, the $R$ parity violating couplings might be non-zero but extremely small, leading to a distinctly different phenomenology.

Assuming $R$ parity conservation, the lightest supersymmetric particle is stable.  In this case, it must be electrically neutral, presumably some
linear combination of the neutral higgsinos and gauginos.  The existence of this stable particle implies that production of supersymmetric
partners in accelerators would be associated with missing energy.  Particularly remarkable is that this particle is a dark matter candidate,
produced in roughly the right quantities in a hot early universe to account for the observed matter density.  Extensive searches have been undertaken and are currently under way for such particles, both through their collision with detectors deep underground (``direct detection") and their annihilations in the cosmos (``indirect detection").

Another striking feature of the MSSM is the unification of the gauge couplings.  For a theory with the particle content of the MSSM, assuming that all of the new particles have masses of order $1$ TeV, one obtains unification of the known gauge couplings, with reasonable accuracy, at a scale $M_{gut} = 2 \times 10^{16}$ GeV, corresponding to a unified coupling $\alpha_{gut} \approx 1/30$.  It is remarkable that these two predictions are outcomes of other requirements, and that they are consequences of symmetry.

Even before the dedicated searches for supersymmetric particles conducted at LEP, the Tevatron, and most recently the LHC, there were significant constraints on these parameters.  The absence of flavor-changing neutral currents in the weak interactions of hadrons requires, in particular, a significant degree of degeneracy (or alignment\cite{nirseiberg}) in the spectrum.  This might be natural, since in the limit of exact degeneracy,
the soft parameters exhibit a significant degree of symmetry.  This requires
special features in the microscopic theory, achieved to date only in models of gauge mediation\cite{Giudice:1998bp} and ``mass matrix models"\cite{massmatrixmodels}.  

\subsection{Supersymmetry:  Detailed Considerations of Naturalness}

The MSSM has provided a paradigm for experimental searches for supersymmetry as well as theoretical efforts to construct a compelling picture of dynamical supersymmetry breaking.  Notions of naturalness lead to certain expectations for the soft-breaking parameters.  

We have mentioned the problems of flavor.  For this there are plausible solutions.    A much more serious challenge to the naturalness principle is the mass of the Higgs particle itself.  Classically within the MSSM, there is a bound on the mass of the lightest Higgs:
\beq
m_H < M_Z
\eeq
\begin{figure}
\includegraphics[width=8cm]{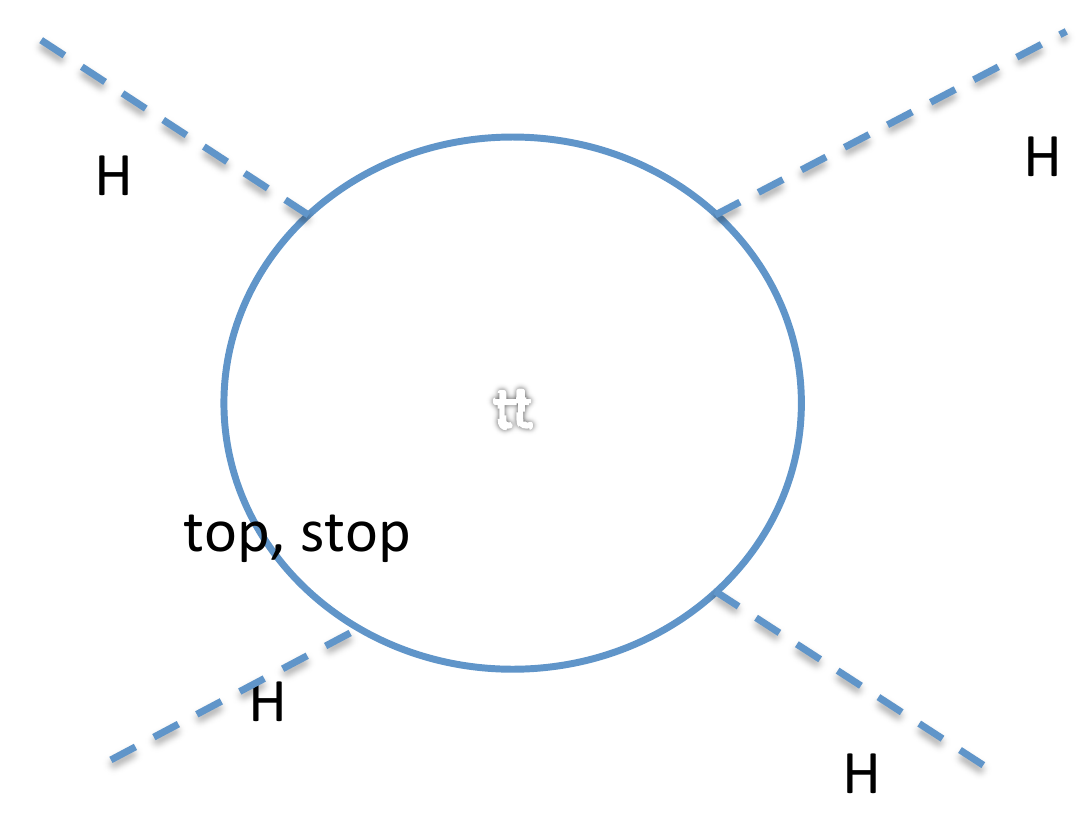}
\caption{Top, stop corrections to the Higgs quartic coupling in the MSSM.}
\end{figure}
This arises because supersymmetry strongly constrains the quartic couplings of the Higgs fields, and these are related to the gauge couplings.
It turns out, however, that due to the top quark, there are significant radiative corrections to the Higgs potential\cite{haberhempfling}.  The diagram of fig. 3, in particular, gives a correction behaving roughly as
\beq
\delta \lambda = {12 y_t^4 \over 16 \pi^2} \log{\tilde m_t \over m_t}
\eeq
More detailed studies yield results of the sort shown in fig. 4.
\begin{figure}
\includegraphics[width=9cm]{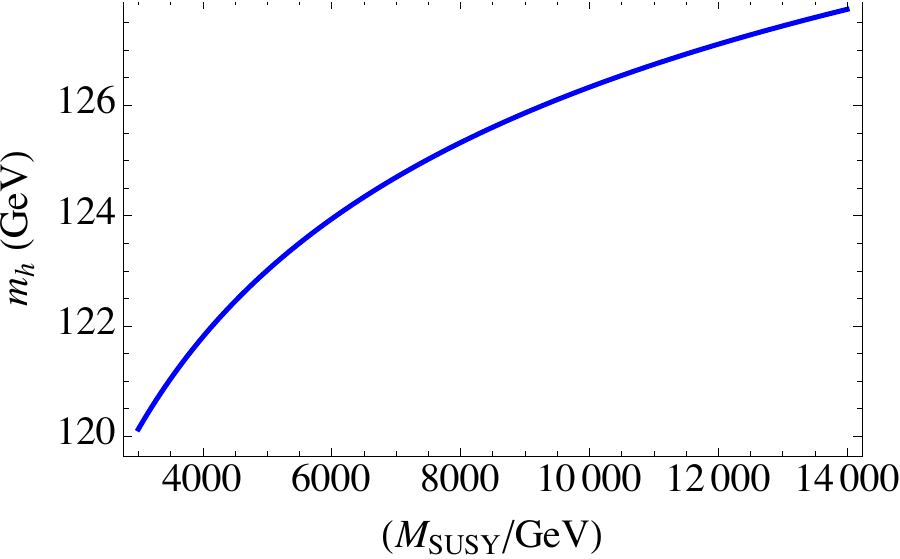}
\caption{Higgs mass as a function of the stop mass for large tan $\beta$, small value of the $A$ parameter.  Includes only leading log corrections.  More complete and detailed
results can be found, e.g,, in \cite{draperhiggs}.}
\end{figure}  

From this figure, we see that, at least within the MSSM, the mass of the recently discovered Higgs particle, $m_H \approx 125$ GeV, requires that the stop be quite heavy, $8$ TeV or more (alternatively one can tune the so-called ``A-Parameter" and obtain a lower stop mass).  This has troubling implications for naturalness.   If we substitute $8$ TeV on the right hand side of Eqn. \ref{stopcontribution} for $\tilde m_t$, and take the ultraviolet cutoff to be, say, $10^{16}$ GeV, then we have that the correction to the Higgs mass parameter is of order $10^4~M_Z^2$, a tuning of parameters of a part in $10^4$. 

Modifying the structure of the MSSM can help with this.  If one adds a gauge singlet field, one can increase the quartic coupling to some degree, and obtain the observed Higgs mass with significantly less tuning (though still appreciable tuning\cite{hallruderman}).

The experimental programs at LEP, the Tevatron and the LHC have provided significant further constraints.    For a broad swath of the parameter space, independent of the arguments about the Higgs mass, squark and gaugino masses are now known to be greater than 1 TeV through much of the parameter space.  This corresponds to tuning at greater than the $1\%$ level, independent of what physics might account for the large mass of the Higgs boson.

\section{The Cosmological Constant and Inflation}

Within the framework of known physics, there is a far more serious violation of naturalness which we have not yet confronted:  the size of the dark energy or cosmological constant (cc).  A cosmological constant is a dimension zero term in the effective action, even more problematic
than the dimension two Higgs mass term:
\beq
{\cal L}_{\Lambda} = \int d^4 x \sqrt{g} \Lambda.
\eeq
Assuming that the observed dark energy is a cosmological constant, we have
$\Lambda \approx 10^{-47} ~{\rm GeV}^4.$
This is an extremely small number in particle physics units; absent any general principle, one might have expected $\Lambda \approx M_p^4$, roughly $120$ orders of magnitude larger.  As for the Higgs mass problem, this estimate is reinforced by a simple-minded calculation\footnote{As stressed by Banks and Fischler, the arguments below are firmly rooted in the framework of effective field theory, and there are good reasons to be skeptical that they apply in a more complete theory of quantum gravity\cite{bankseft}.Frameworks in which
a small c.c. might be more natural have been discussed in \cite{banksfischler,thomascc} and elsewhere.}.  In a quantum field theory, even if the vacuum energy vanishes classically, there is a quantum contribution to the energy, which is just a sum of the zero point energy for bosons and the energy of the filled Dirac sea for fermions,
\beq
\Lambda = \sum_{helicities} (-1)^F {1 \over 2} \int {d^3 k \over (2 \pi)^3} \sqrt{k^2 + m^2}.
\eeq
Here $(-1)^F$ is $+1$ for bosons and $-1$ for fermions.  Each term in the sum is quartically divergent.  Taking $M_p$ as the cutoff yields the naive estimate.

In the case of supersymmetric theories, things are somewhat better.  The number of bosonic and fermionic degrees of freedom is the same, and the leading divergence cancels.  But one gets a result proportional to the fourth power of the supersymmetry-breaking scale.  Even for the lowest conceivable SUSY breaking scale (TeV), this is many orders of magnitude larger than the observed dark energy.

In fact, there is no proposal to understand the small value of the dark energy in 'thooftian terms; General Relativity simply does not become more symmetric in the limit $\Lambda \rightarrow 0$.  Calculations in string theory, the only framework we have where the dark energy
may be calculable, are consistent with expectations based on dimensional analysis (\cite{rohm}). 

The value of the c.c. is remarkable in another way.  While small in particle physics units, it is substantial in units relevant to the present cosmological epoch; indeed, the c.c. has just become important ``recently" (the past few billion years), and it will dominate the energy density ``forever".  One could imagine that some dynamics couples the c.c. and the density of dark matter, for example, but no such connection has been uncovered.  Instead Weinberg\cite{weinbergcc}, following a suggestion of Banks\cite{bankscc} and Linde\cite{lindecc} proposed an explanation of a different type.  He imagined, in essence, that the observed universe is part of a larger structure, subsequently dubbed a ``multiverse", in which the c.c. can take a range of values, essentially randomly distributed.  If one could take an inventory of this mumultiverse, one would find that only in some regions are there observers.  This criterion, know as the {\it anthropic principle}, is much like arguing that observers (e.g. humans) are only found in a very tiny fraction of the volume of the universe, on the surfaces of planets with liquid water.

At a minimum, Weinberg argued, a universe supporting observers should contain galaxies.  In our universe, galaxies formed about 1 billion years or so after the big bang; we understand this as the time required for small primordial density fluctuations (presumably formed during an epoch of inflation) to grow and become non-linear.  If the c.c. were so large that it dominated the energy density 1 billion years after the big bang, structure would not form.  

An additional, crucial element of the argument relies precisely on the fact that the c.c. is unnatural:  there is nothing more symmetric or otherwise special about a lagrangian with vanishing $\Lambda$, so that it is reasonable to expect that the probability of finding one or another value of $\Lambda$ near $\Lambda =0$ is uniform.  So, in particular, one is likely to find the {\it largest} value of $\Lambda$ consistent with the anthropic requirement above.  This is somewhat larger than the dark energy which was subsequently discovered.  More refined versions of the argument\cite{aprioricc} lead to values closer to the observed value.

One may or may not be troubled by entertaining the possibility that anthropic considerations determine the laws of physics, and one can debate how significant is the success of predicting, at least at a rough order of magnitude level, the c.c. 
Perhaps a more compelling concern raised by such considerations is simply:  do there exist physical theories in which such a possibility is realized.  The number of possible configurations which must be surveyed is enormous; given the small value of the c.c. in typical particle physics units, one might imagine that there should be at least $10^{120}$ such states.  A number of researchers have put forward scenarios in which such a ``landscape" of possibilities, usually thought of as (metastable) vacua of some underlying theory, might arise\cite{irrationalaxion,boussopolchinski,kklt}.  In string theory with some compactified dimensions, in particular, there are many types of quantized flux (analogous to magnetic flux in QED) which can take many values, giving the potential for vast numbers of possible states.  In each of these vacua, the degrees of freedom and the parameters of the lagrangian will take different values.  If there are enough such states, the parameters will be densely distributed. The existence of such a {\it landscape} or {\it discretuum} of vacua remains conjectural, however.

The success, to the degree that it may be counted as one, of anthropic considerations for the c.c. raises the possibility (concern(?)) that such considerations might govern other features of our Standard Models, most notably the Higgs mass.  Indeed, this
mass is not nearly as severely tuned as the c.c.  Moreover, it is plausible that the TeV scale is anthropically selected.  If the Higgs mass-squared were much larger than it is, one would either  electroweak symmetry would be unbroken, or it would be broken and and the $W$'s and $Z$  {\it extremely} heavy.  In either case, life would likely be impossible.  If stars existed at all, their properties would be quite different than those in our universe, affecting important quantities like the abundance of heavy
elements. \footnote{For a recent, wide-ranging discussion of these issues, with extensive references, see \cite{halletal}.}

So it is conceivable that the value of the Higgs mass is selected by anthropic considerations from a landscape of possibilities.  If so, the naturalness principle {\it might} not be operative, and the value of the electroweak scale might not have any additional consequences for low energy physics.

Other aspects of cosmology raise serious questions of naturalness as well.  {\it Inflation}, the proposal that the universe went through a period of extremely rapid expansion early in its history, has received extensive experimental support in the past two decades from studies of the Cosmic Microwave Radiation Background\cite{planckinflation}.  Inflation explains the homogeneity and flatness of the universe, and the structure we observe about us,  but existing models of the phenomenon suffer from problems of fine tuning in varying degrees .  It is plausible that anthropic considerations might play some role here as well.

\section{Other Arenas for Questions of Naturalness}

We have already mentioned another puzzling number in the Standard Model:  the small value of $\theta_{qcd}$.  Interestingly, this is a question which is not likely to be solved anthropically\cite{donoghue,banksdinegorbatov}.  Provided $\theta$ is less than some moderately small number (certainly not more than $0.01$), nothing changes qualitatively in the strong interactions; indeed, the dependence on $\theta$ of nuclear reaction rates, for example, is very weak\cite{ubaldi}.  

Solutions which are compatible with notions of naturalness have been put forward.  They rely, ultimately, on the fact that QCD. considered in isolation, becomes more symmetric in the limit $\theta \rightarrow 0$.  One possibility is that the mass of the $u$ quark is very small; In the limit $m_u \rightarrow 0$, $\theta$ is unobservable and $CP$ is conserved in the strong interactions. $d_n$ is smaller than the experimental limit provided
\beq
{m_u \over m_d} < 10^{-10}.
\eeq
 The main question is whether a small $u$ quark mass is compatible with facts of the strong interactions.  There has been debate about this question through the years\cite{georgimcarthur,kaplanmanohar,banksnirseiberg}, but lattice gauge theory calculations appear to conclusively rule out this possibility\cite{review1,review2,BMW}.

A second proposal is the ``axion", a pseudogoldstone particle associated with an approximate global ``Peccei-Quinn" symmetry.  This field would couple to $F \tilde F$.  If its potential only arises through this coupling, it has a minimum near the origin, where the theory conserves CP.  The approximate symmetry which accounts for this must be an extremely good symmetry.  The axion mass is of order 
\beq
m_a^2 \approx {m_\pi^2 f_\pi^2 \over f_a^2}
\eeq
which, for $f_a$'s of order $10^{11}$ GeV or larger (as required from astrophysical considerations), is extremely small.  For a range
of parameters (and depending on assumptions about early cosmic history) the axion can be the dark matter.

As a consequence of the small axion mass, tiny, CP-asymmetric couplings can give rise to an unacceptably large $\theta$.
A number of proposals have been put forward to achieve a Peccei=Quinn symmetry of sufficient
quality, the most compelling coming from string theory\cite{wittenaxion,bobkovraby,dineaxions}.  In an interesting range of its parameter space, this particle can play the role of dark matter (which does raise the possibility that there might be some sort of anthropic selection for axions and hence small $\theta$\cite{dineetalquality}).

A third proposal is that CP is conserved in the underlying theory, and spontaneously broken in a way that generates an order one KM angle
with a tiny $\theta$.  Models for such a phenomenon have been put forward in \cite{nelsoncp,barrcp}.  There are, however, many difficulties in assuring
that $\theta$ remains small when higher dimension operators and quantum corrections are taken int account.  In a landscape framework, as we will shortly discuss, while CP is, indeed, conserved in the underlying theory, CP conserving ground states (i.e. states where the ``bare" $\theta$ might be expected to be zero) are likely to be very rare.  At least at this time, then, it appears that the axion is the most plausible solution of the
strong CP problem.

Once one has admitted the possibility of anthropic selection, one is forced to contemplate its relevance even for quantities which are naturally
small.  One might well imagine that anthropic considerations could play a role in determining the masses of the $u,d$ quarks and the electron,
though there possible relevance for heavier quarks and leptons is not obvious.   

\section{Model Landscapes}



We have already mentioned that compactification of string theory with fluxes provides a model for how a landscape might arise.  In interesting constructions, the number of possible flux types is often large (of order $100$'s or more), and these fluxes can range over many discrete values.  For each choice of flux, there may be many stationary points of the effective action.  In this way, one can build up an exponentially large number of states; this is a setting for Weinberg's solution of the cosmological constant problem.

There are many challenges to establishing the existence of a discretuum.
Before turning on fluxes, in the classical approximation, string vacua exhibit large, continuous degeneracies.  Associated with these degenearcies are large numbers of scalar fields, called moduli, without potentials.  Turning on fluxes often gives potentials for many of these fields, with stable minima.  But, again at the classical level, there are invariably some massless fields left over.  It is plausible that some or all of these remaining fields are stabilized by non-perturbative effects.  Scenarios were put forward in \cite{kklt}.  These authors argued for the existence of isolated vacua with supersymmetry or approximate supersymmetry.  Actually constructing such vacua in
a consistent manner is challenging; it is debatable, for example, whether there is ever a small parameter which allows systematic study.

Assuming the existence of a landscape, the interesting question is to understand the statistics of these states.    One might hope, given a knowledge of the distribution of parameters and
some observational or anthropic constraints, to establish that, say low energy supersymmetry is or is not likely; indeed, as we will discuss
further in section \ref{landscapenaturalness}, this would provide a quite concrete realization of notions
of naturalness.  There has been some effort to understand such statistics\cite{douglasdenef1}. Plausible arguments have been put forward, for example, that
\begin{enumerate}
\item  Among non-supersymmetric stationary points, only a very tiny fraction are metastable.  This suggests, but hardly proves, that some degree of supersymmetry might be an outcome\cite{mcalister}.
\item  Among remaining non-supersymmetric states, with small cc, the vast majority are short-lived\cite{dinefestucciamorisse,greeneweinberg}.
\item  Among supersymmetric states, if supersymmetry is not dynamically broken, high scales of supersymmetry
breaking are favored\cite{douglasdenef2,dgt,dinebranches}.  With dynamical breaking, lower scales may be favored.
\item  As we will discuss further below, states exhibiting certain types of (ordinary) symmetries are rare.
\end{enumerate}
Even lacking a completely reliable model, assuming the existence of a landscape, many of these features would seem robust.  They rely on quite minimal assumptions about the features of low energy effective actions and distributions of lagrangian parameters.

\section{Naturalness in a Landscape Framework}
\label{landscapenaturalness}

We have presented the rather bleak prospect that certain parameters of the Standard Model, such as the Higgs mass, are completely determined by anthropic considerations, and considerations of naturalness, and with them interesting possibilities for new TeV scale
degrees of freedom and new symmetries, play no role.  But there are intermediate possibilities, and this question should be considered with greater care.

Indeed, a landscape in some sense provides an ideal setting to consider questions of naturalness and to understand how it might emerge, sometimes or always, as a governing principle.  Weinberg's cosmological constant argument relies crucially on the assumption that there is nothing special, at a fundamental level, about the point where $\Lambda =0$.  For the Higgs, things might be different if nature is approximately supersymmetric.  Indeed, in studies of model landscapes\cite{douglasdenef1,douglasdenef2,dgt,dinebranches}, several branches have been identified which differ in the nature of the realization of supersymmetry:
\begin{enumerate}
\item  Non-supersymmetric branch.  Here the distribution of Higgs mass-squared is roughly uniform.  The cost of having a Higgs mass $m_H$ is $m_H^2/M_p^2$.
\item  Supersymmetric branch where the breaking of supersymmetry is non-dynamical:  Here supersymmetry, despite the cancellation of quadratic divergences, does not help; the fraction of states with larger breaking of supersymmetry, $F$, grows as a large power of $F$.  So operationally, this branch is like branch (1).
\item  Supersymmetric branch with dynamically broken supersymmetry (in the sense that the supersymmetry breaking scale behaves as $e^{-{8 \pi^2 \over b g^2}}$).  Here the number of states with small Higgs mass and small c.c. is the same per decade as a function of the supersymmetry breaking scale.  Without introducing other considerations (perhaps the density of dark matter) there is no preference for low scale supersymmetry breaking.  Conceivably such other considerations would favor a scale more like $8$ TeV than $1$ TeV.
\item  Supersymmetric branch favoring low scale supersymmetry breaking:  on this branch other quantities (the value of the superpotential and the $\mu$ parameter) are dynamically determined as well (corresponding to dynamical breaking of so-called $R$ symmetries).  Here the lowest possible scale of supersymmetry breaking is favored.  General arguments can be put forward suggesting that there are far less states on this branch than the third one.
\end{enumerate}

\subsection{How Natural Are Symmetries?}

In a landscape framework, one can revisit the question of symmetries themselves.  The symmetries we have in mind are discrete symmetries, global continuous symmetries, gauge symmetries and supersymmetry.  't Hooft's naturalness principle assumes  that symmetries, themselves, are special or singled out.   Of these various types of symmetries, global continuous symmetries are not a feature of quantum gravity theories (in string theories, this is often a theorem\cite{banksdixonsymmetries}).  Gauge symmetries appear common in string theories, as does supersymmetry.  Discrete symmetries appear frequently as well.   It is these latter symmetries which are of particular interest.  They might account for the stability of the proton in supersymmetric theories, and the smallness of the Yukawa couplings of the standard model, and it is usually assumed, in building particle physics models, that they are somehow singled out.  Yet,  in the flux landscapes which have been studied states (vacua) with symmetries would appear to be quite rare\cite{dinesun}.  

To understand this, we can ask how symmetries arise when one compactifies a theory on some compact space.  In many solutions of string theory the compact space exhibits discrete symmetries.  These are typically subgroups of the original rotational symmetry of the higher dimensional space.  These symmetries translate into conventional discrete symmetries of the field theory which describes the system at low energies.  Now imagine turning on fluxes.  Typical fluxes will transform under these rotations; as a result, the low energy theory {\it does not} exhibit the symmetry.  Recall that in the flux landscape, the large number of states results from the large number of possible fluxes.  If most of the fluxes are not invariant under the symmetry -- the typical situation -- then at best an exponentially small fraction of the states will exhibit the symmetry.  These considerations apply to the sorts of discrete symmetries we might invoke to explain proton stability, and also to CP.

There may be other considerations (cosmological?) which  would favor symmetric states\cite{Dine:2008jx}.  But this simple observation  {\it calls into question the basic assumptions of 't Hooft's naturalness criterion}.

Interestingly, supersymmetry {\it might} function differently.  Another issue in a landscape is {\it stability}; a state of small cc, similar to our own, will be surrounded by vast numbers of states with negative c.c.  It is necessary that the lifetime for decay of the state to {\it every one} of its neighbors be extremely small\cite{greeneweinberg,dinefestucciamorisse}.  It turns out that the simplest way to account for such stability is approximate supersymmetry of the state.  In the limit of exact supersymmetry, in fact, the symmetry insures exact stability; if the breaking is small, the lifetime of the state becomes exponentially long as the breaking scale becomes small.\cite{dinefestucciamorisse}.

\section{Conclusions:  Naturalness as a Guide}  

it is still possible that nature is ``natural", in the sense of 't Hooft.  Future runs of the LHC might provide evidence for supersymmetry, warped extra dimensions, or some variant of technicolor.  But the current experimental situation raises the unsettling possibility that naturalness may not be a good guiding principle.  Indeed, naturalness is in tension with another principle:  simplicity.  Simplicity has a technical meaning:  the simplest theory is the one with the smallest
number of degrees of freedom consistent with known facts.  Contrast, for example, the minimal Standard Model, with its single Higgs doublet, with supersymmetric theories, with their many additional fields and couplings.  So far, the experimental evidence suggests that simplicity is winning.  The observed Higgs mass is in tension with expectations from supersymmetric theories, but also technicolor and other proposals.

On the other hand, the main alternative to natural theories (apart from the possibility that {\it extreme} fine tuning is simply a fact) is the landscape or multiverse.  In such a situation, our neighborhood in the universe might be simple, but the underlying structure is unimaginably complex.  We have seen, however, that this idea has at least one major success:  the prediction of the dark energy. It provides a plausible picture for other (but not necessarily all) tuned quantities.  


Why might we subscribe to a naturalness principle?  After all, if the universe is described by a single theory, with a single set of degrees of freedom and a single lagrangian with fixed parameters, the question of fine tuning is metaphysical; things are the way we are, and it is not clear why we should be troubled the value of some parameter or other.  We might hope that if things are unique
The landscape has the potential to make the question concrete. If we simply ask:  where are the most states consistent with nature as observed
(small c.c., large hierarchy), we have seen that model landscapes may prefer, for example, no supersymmetry or very high scale of supersymmetry
breaking.  Conventional symmetries, such as discrete symmetries (including CP) would seem likely rare.  On the other hand, we have given some arguments that supersymmetric states might be common, and that classes of these would favor supersymmetry in the conventional way.

It is possible that the next round of LHC experiments will discover evidence for supersymmetry, large extra dimensions, or totally unanticipated phenomena which will restore our confidence in the notion of naturalness.

\section*{ACKNOWLEDGMENTS}
This work was supported by the U.S. Department of Energy grant number DE-FG02-04ER41286.   I thank many colleagues for discussions of these issues,
particularly Tom Banks, Savas Dimopoulos, Patrick Draper, and Nima Arkani-Hamed.

 \newpage

\bibliographystyle{unsrt}
\bibliography{naturalness_under_stress.bbl}{}
\end{document}